\documentclass[Afour,times,twocolumn]{sagej} 

\def\volumeyear{2014}
\def\journalname{-}
\def\volumenumber{-}
\def\issuenumber{-}

\usepackage{apacite}
\let\cite\shortcite

\usepackage{amsmath}
\usepackage{pifont}
\def\*#1*\ {}
\usepackage{float}
\usepackage{array}
\usepackage{tikz}
\usetikzlibrary{shapes.geometric}
\usepackage{tikz}
\usetikzlibrary{arrows}
\usepackage{tkz-berge}

\usepackage{calc}
\usetikzlibrary{calc}
\usepackage{url}

\usepackage{graphicx}
\usepackage{caption}
\usepackage{subcaption}
\usepackage{balance}
\usepackage{picins}

\usepackage{xcolor}
\definecolor{myorange}{rgb}{0.9882352941176471, 0.5529411764705883, 0.3843137254901961}
\definecolor{mygreen}{rgb}{0.4, 0.7607843137254902, 0.6470588235294118}

\newcommand{\matr}[1]{\mathbf{#1}}

\affiliation{*Electronics and Information Systems \\ Ghent University}

\begin{document}

\runninghead{Degrave}
\title{Resolving multi-proxy transitive vote delegation}
\author{Jonas Degrave\affilnum{1}}
\affiliation{\affilnum{1}Electronics and Information Systems, Ghent University, Belgium}
\corrauth{Jonas Degrave, Sint-Pietersnieuwstraat 41, 9000 Gent, Belgium}
\email{Jonas.Degrave@UGent.be}

\begin{abstract}
Solving a delegation graph for transitive votes is already a non-trivial task for many programmers. When extending the current main paradigm, where each voter can only appoint a single transitive delegation, to a system where each vote can be separated over multiple delegations, solving the delegation graph becomes even harder. This article presents a solution of an example graph, and a non-formal proof of why this algorithm works.
\end{abstract}
\maketitle

\section*{Introduction}

In the area of voting systems, there is a growing interest into the topic of liquid democracy, in which votes can be delegated to others in order to deal with the complexity of the topic at hand.
In these voting systems, many currently use a single proxy system, in which you delegate your vote to one other user. This other user can in its turn further delegate his and your vote in a transitive chain. However, when the user who receives the delegated vote decides not to use it (possibly unintentionally), the vote goes to waste.

However, there is no mathematical need for a user to only delegate to a single other user. In this paper, I propose a multi-proxy delegation system. This system fulfills the following requirements:
\begin{itemize}
\item Every user has one vote, which can be delegated.
\item The delegation is transitive, meaning that it can be further delegated.
\item This vote is distributed equally between all the user's direct delegates who actually vote, potentially through further delegations.
\end{itemize}

This approach can be used in addition to vote counting systems, as long as each vote in the counting system can be weighted with the total number of votes each user received.

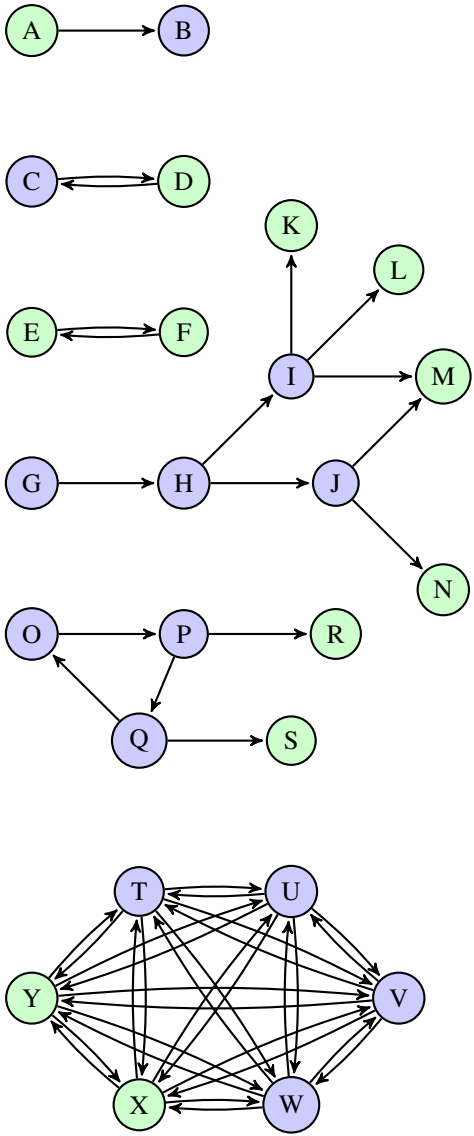
\begin{figure}[htpb]
    \begin{tikzpicture}[->,>=stealth',shorten >=1pt,auto,node distance=2cm,
      thick,main node/.style={circle,fill=blue!20,draw},vote node/.style={circle,fill=green!20,draw}]
      \node[vote node] (A) {A};
      \node[main node] (B) [right of=A] {B};
      \node[main node] (C) [below of=A] {C};
      \node[vote node] (D) [right of=C] {D};
      \node[vote node] (E) [below of=C] {E};
      \node[vote node] (F) [right of=E] {F};
      \node[main node] (G) [below of=E] {G};
      \node[main node] (H) [right of=G] {H};
      \node[main node] (I) [above right of=H] {I};
      \node[main node] (J) [right of=H] {J};
      \node[vote node] (K) [above of=I] {K};
      \node[vote node] (L) [above right of=I] {L};
      \node[vote node] (M) [right of=I] {M};
      \node[vote node] (N) [below right of=J] {N};
      
      \node[main node] (O) [below of=G] {O};
      \node[main node] (P) [right of=O] {P};
      \node[main node] (Q) [below right of=O] {Q};
      \node[vote node] (R) [right of=P] {R};
      \node[vote node] (S) [right of=Q] {S};

      \node[main node] (T) [below of=Q] {T};
      \node[main node] (U) [right of=T] {U};
      \node[main node] (V) [below right of=U] {V};
      \node[main node] (W) [below left of=V] {W};
      \node[vote node] (X) [left of=W] {X};
      \node[vote node] (Y) [above left of=X] {Y};
    
      \path[bend angle=5,every node/.style={font=\sffamily\small,}]
        (A) edge             node [left]  {} (B)
        (C) edge [bend left] node [left]  {} (D)
        (D) edge [bend left] node [right] {} (C)
        (E) edge [bend left] node [left]  {} (F)
        (F) edge [bend left] node [right] {} (E)
        (G) edge             node [right] {} (H)
        (H) edge             node [right] {} (I)
            edge             node [right] {} (J)
        (I) edge             node [right] {} (K)
            edge             node [right] {} (L)
            edge             node [right] {} (M)
        (J) edge             node [right] {} (N)
            edge             node [right] {} (M)
        (O) edge             node [right] {} (P)
        (P) edge             node [right] {} (Q)
            edge             node [right] {} (R)
        (Q) edge             node [right] {} (O)
            edge             node [right] {} (S)
            
        (T) edge [bend left] node {} (U) edge [bend left] node {} (V) edge [bend left] node {} (W) edge [bend left] node {} (X) edge [bend left] node {} (Y)
        (U) edge [bend left] node {} (V) edge [bend left] node {} (W) edge [bend left] node {} (X) edge [bend left] node {} (Y) edge [bend left] node {} (T)
        (V) edge [bend left] node {} (W) edge [bend left] node {} (X) edge [bend left] node {} (Y) edge [bend left] node {} (T) edge [bend left] node {} (U)
        (W) edge [bend left] node {} (X) edge [bend left] node {} (Y) edge [bend left] node {} (T) edge [bend left] node {} (U) edge [bend left] node {} (V)
        (X) edge [bend left] node {} (Y) edge [bend left] node {} (T) edge [bend left] node {} (U) edge [bend left] node {} (V) edge [bend left] node {} (W)
        (Y) edge [bend left] node {} (T) edge [bend left] node {} (U) edge [bend left] node {} (V) edge [bend left] node {} (W) edge [bend left] node {} (X)
        ;
    \end{tikzpicture}
    \caption{The delegation graph for 25 people we solve as an example in this paper. These people either vote themselves, and are shown as green nodes, or they did not vote, and are shown as blue nodes. Every edge from A to B represents a delegation from person A to person B.}
    \label{fig:graph:original}
\end{figure}

\section*{Preparing the example}

Suppose we have the following delegation graph for 25 people, as presented in Figure~\ref{fig:graph:original}. These people either vote themselves, and are shown as green nodes, or they did not vote, and are shown as blue nodes. We chose this example to show all relevant difficulties encountered in solving the delegation graph.

In order to tackle the problem using a flow graph, there are still multiple problems which need fixing. Therefore, we need the following initial steps:
\begin{itemize}
\item Remove all edges starting from people who vote themselves.
\item Descend the tree starting from the people who vote themselves, and collect all nodes encountered this way. Remove all nodes not encountered from this descend, as their vote cannot be cast in a meaningful way.
\end{itemize}

\begin{figure}[htpb]
    \begin{tikzpicture}[->,>=stealth',shorten >=1pt,auto,node distance=2cm,
      thick,main node/.style={circle,fill=blue!20,draw},vote node/.style={circle,fill=green!20,draw}]
      \node[vote node] (A) {$A$};
      \node[main node] (C) [below of=A] {$C$};
      \node[vote node] (D) [right of=C] {$D$};
      \node[vote node] (E) [below of=C] {$E$};
      \node[vote node] (F) [right of=E] {$F$};
      \node[main node] (G) [below of=E] {$G$};
      \node[main node] (H) [right of=G] {$H$};
      \node[main node] (I) [above right of=H] {$I$};
      \node[main node] (J) [right of=H] {$J$};
      \node[vote node] (K) [above of=I] {$K$};
      \node[vote node] (L) [above right of=I] {$L$};
      \node[vote node] (M) [right of=I] {$M$};
      \node[vote node] (N) [below right of=J] {$N$};
      
      \node[main node] (O) [below of=G] {$O$};
      \node[main node] (P) [right of=O] {$P$};
      \node[main node] (Q) [below right of=O] {$Q$};
      \node[vote node] (R) [right of=P] {$R$};
      \node[vote node] (S) [right of=Q] {$S$};

      \node[main node] (T) [below of=Q] {$T$};
      \node[main node] (U) [right of=T] {$U$};
      \node[main node] (V) [below right of=U] {$V$};
      \node[main node] (W) [below left of=V] {$W$};
      \node[vote node] (X) [left of=W] {$X$};
      \node[vote node] (Y) [above left of=X] {$Y$};
    
      \path[bend angle=5,every node/.style={font=\sffamily\small,}]
        (C) edge [bend left] node [left]  {} (D)
        (G) edge             node [right] {} (H)
        (H) edge             node [right] {} (I)
            edge             node [right] {} (J)
        (I) edge             node [right] {} (K)
            edge             node [right] {} (L)
            edge             node [right] {} (M)
        (J) edge             node [right] {} (N)
            edge             node [right] {} (M)
        (O) edge             node [right] {} (P)
        (P) edge             node [right] {} (Q)
            edge             node [right] {} (R)
        (Q) edge             node [right] {} (O)
            edge             node [right] {} (S)
            
        (T) edge [bend left] node {} (U) edge [bend left] node {} (V) edge [bend left] node {} (W) edge [bend left] node {} (X) edge [bend left] node {} (Y)
        (U) edge [bend left] node {} (V) edge [bend left] node {} (W) edge [bend left] node {} (X) edge [bend left] node {} (Y) edge [bend left] node {} (T)
        (V) edge [bend left] node {} (W) edge [bend left] node {} (X) edge [bend left] node {} (Y) edge [bend left] node {} (T) edge [bend left] node {} (U)
        (W) edge [bend left] node {} (X) edge [bend left] node {} (Y) edge [bend left] node {} (T) edge [bend left] node {} (U) edge [bend left] node {} (V)
        ;
    \end{tikzpicture}
    \caption{The simplified delegation graph.}
    \label{fig:graph:simplified}
\end{figure}
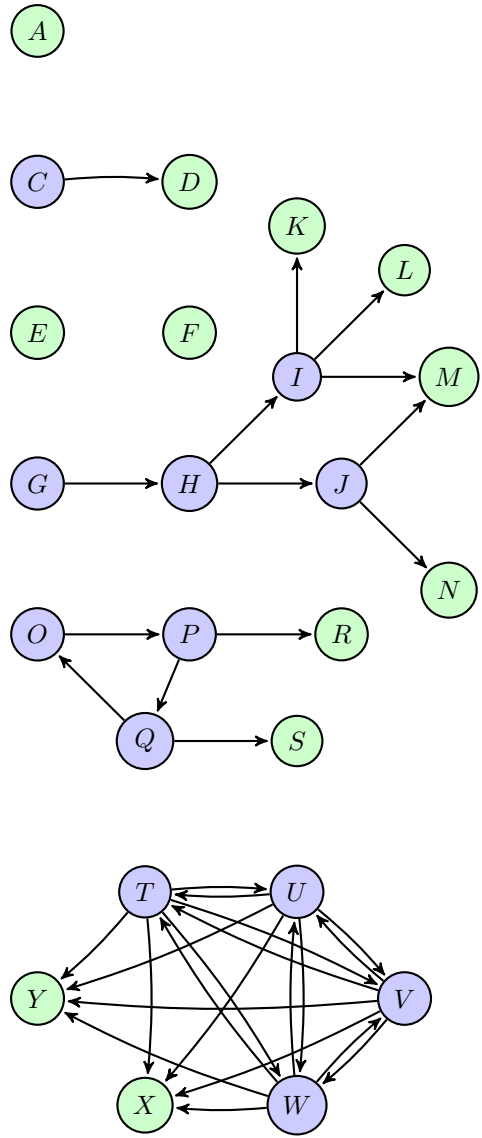

After this process, we end up with the graph depicted in Figure~\ref{fig:graph:simplified}.

\section*{Solving the delegation graph: the first method}
In this section, I will introduce a first approach to solving this delegation graph, based on solving a system of linear equations. As you can see, $D$ receives one vote from $C$ in addition to the vote he already had. Therefore:
\begin{equation*}
D = 1 + C.
\end{equation*}
M started with a vote, and receives one third of a vote from I, and half of a vote from J:
\begin{equation*}
M = 1 + \frac{1}{3} I + \frac{1}{2} J
\end{equation*}
If we do this for all votes, we end up with the following set of linear equations.
{\allowdisplaybreaks
\begin{align*}
A &= 1 &\\
C &= 1 &\\ 
D &= 1 + C &\\
E &= 1 &\\
F &= 1 &\\
G &= 1 &\\
H &= 1 + G &\\
I &= 1 + \frac{1}{2} H &\\
J &= 1 + \frac{1}{2} H &\\
K &= 1 + \frac{1}{3} I &\\
L &= 1 + \frac{1}{3} I &\\
M &= 1 + \frac{1}{3} I + \frac{1}{2} J &\\
N &= 1 + \frac{1}{2} J &\\
O &= 1 + \frac{1}{2} Q &\\
P &= 1 + O &\\
Q &= 1 + \frac{1}{2} P &\\
R &= 1 + \frac{1}{2} P &\\
S &= 1 + \frac{1}{2} Q &\\
T &= 1 + \frac{1}{5} U + \frac{1}{5} V + \frac{1}{5} W &\\
U &= 1 + \frac{1}{5} T + \frac{1}{5} V + \frac{1}{5} W &\\
V &= 1 + \frac{1}{5} T + \frac{1}{5} U + \frac{1}{5} W &\\
W &= 1 + \frac{1}{5} U + \frac{1}{5} V + \frac{1}{5} W &\\
X &= 1 + \frac{1}{5} T + \frac{1}{5} U + \frac{1}{5} V + \frac{1}{5} W &\\
Y &= 1 + \frac{1}{5} T + \frac{1}{5} U + \frac{1}{5} V + \frac{1}{5} W &\\
\end{align*}
}
We can solve system this linear system exactly for all variable $A$ through $Y$, by converting it to a a matrix system, as shown in equation~\ref{eq:problem}.

This system can easily be solved by inverting the matrix $\matr{B}$, if it is not singular, and $\matr{B}$ is never singular, since it represents a bijective transformation.
\begin{equation}
  \matr{S} = \matr{B}^{-1}\cdot\matr{J},\label{eq:path1}
\end{equation}
This solution is shown in equation~\ref{eq:solution}.

\section*{Solving the delegation graph: the second method}
A second way to look at this, is by constructing the adjacency matrix $D$ of our directed graph. Now, we know that after zero steps in the delegation chain $\matr{S} = \matr{J}$. After one step in the delegation chain $\matr{S} = \matr{A}.\matr{J}$, after two steps $\matr{S} = \matr{A}^2.\matr{J}$, and so on. Therefore the total number of votes everybody has after an infinite number of delegation steps is
\begin{equation*}
  \matr{S} = \lim_{n \to +\infty}\sum\limits_{i=0}^n \matr{A}^n\cdot\matr{J},
\end{equation*}
and since $\displaystyle\lim_{n \to +\infty}\matr{A}^n=\matr{0}$,
\begin{equation*}
  \matr{S} = \Bigg(\lim_{n \to +\infty}\sum\limits_{i=0}^n \matr{A}^n\Bigg)\cdot\matr{J}.
\end{equation*}
Because this is a Neumann series
\begin{equation}
  \matr{S} = (\matr{I}-\matr{A})^{-1}\cdot\matr{J}. \label{eq:path2}
\end{equation}
Note that $\matr{B}=\matr{I}-\matr{A}$, and hence, the solutions in \ref{eq:path1} and \ref{eq:path2} are identical.

\section*{How to interprete the results}
We could just look at the matrix $\matr{S}$ to find out the total amount of votes the people who actually voted received from the delegation graph. It might be tempting to interpret the elements of $\matr{S}$ of nodes not voting, as the number of votes they would have had, if they would have voted. This is not correct. Take for example the amount of votes of node $O$. The matrix $\matr{S}$ indicates that $O$ would have had 2.333 votes. This is not true, you can easily verify that $O$ actually only would have had 1.75 votes. The reason is that the edge $O \rightarrow P$ is removed if $O$ votes, changing the flow of votes altogether.

But there is more information: the matrix $\matr{B}^{-1}$ contains more useful information about the origin of each vote. If we take for instance the solution to our specific problem shown in equation~\ref{eq:solution}, we can see that each element of $\matr{B}_{ij}$ indicates the contribution of node $j$ to the amount of votes of $i$, if and only if $i$ actually votes. This allows for detailed feedback to each user as to why he voted exactly this, or to why he voted with a certain amount of votes. It is however not possible to trace the exact route of each vote, since it might be -- and often is -- infinitely long.

\balance
\section*{How to implement the algorithm}
While inverting a matrix is a very common operation, this is non-trivial for large networks. A matrix inversion is of the order $O(N^3)$ for speed and $O(N^2)$ with $N$ the rank of the matrix, here the number of nodes. Since our matrix is however most likely sparse, since the expected number of delegations per user $E[D] \ll N$. This allows for more specialized approaches developed especially for sparse matrices, such that the the memory use scales with $O(ND)$ and the speed scales approximately with $O(N^2)$ as well~\cite{li2009fast}. Since usually $E[D]$ is really small, we have used the algorithm provided in the scipy package without notable problems, even though it is slower. It does have the benefit of only needing a limited amount of memory.

\section*{Further extensions}
This approach can easily be extended to allow for other ideas in liquid voting systems, such as decaying votes when the trust chain grows (by making the sum of outgoing edges smaller than 1), or where the user can weigh the distribution of his vote for his proxies autonomously rather than presuming the vote is distributed equally.

\bibliography{must}

\begin{biogs}

\parpic{\includegraphics[width=1in,clip,keepaspectratio]{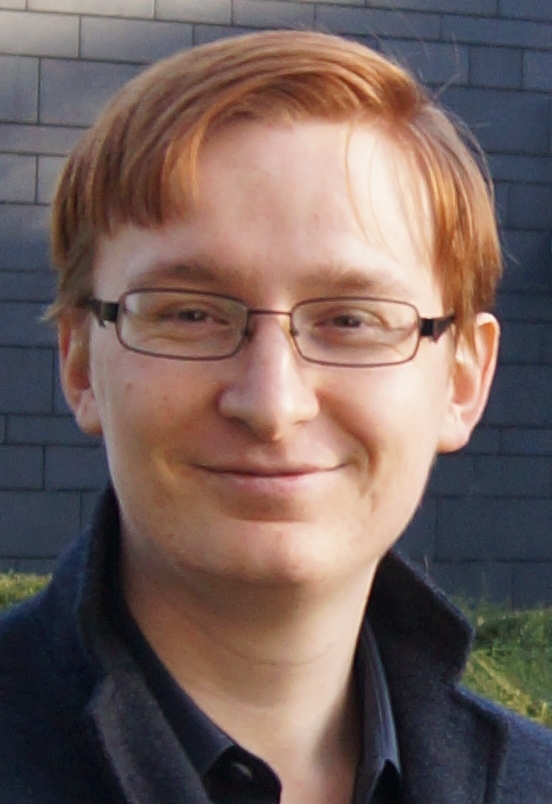}}\noindent \textbf{Jonas Degrave} received a M.Sc. degree in electronics engineering at Ghent University in 2012, where he is currently pursuing a Ph.D. in computer science. His research interests are focused on machine learning and how it can be applied on robotics in order to generate more efficient locomotion.

\end{biogs}

\clearpage

\begin{sideways}
\parbox{\textheight}{
\setlength\arraycolsep{4pt}
\setcounter{MaxMatrixCols}{24}
\begin{equation}
\overbrace{
\begin{pmatrix} 
1 \\
1 \\
1 \\
1 \\
1 \\
1 \\
1 \\
1 \\
1 \\
1 \\
1 \\
1 \\
1 \\
1 \\
1 \\
1 \\
1 \\
1 \\
1 \\
1 \\
1 \\
1 \\
1 \\
1 \\
\end{pmatrix}
}^\matr{J}
=
\overbrace{
\begin{bmatrix} 
1 & & & & & & & & & & & & & & & & & & & & & & &  \\
& 1 & & & & & & & & & & & & & & & & & & & & & & \\
& -1 & 1 & & & & & & & & & & & & & & & & & & & & & \\
& & & 1 & & & & & & & & & & & & & & & & & & & & \\
& & & & 1 & & & & & & & & & & & & & & & & & & & \\
& & & & & 1 & & & & & & & & & & & & & & & & & & \\
& & & & & -1 & 1 & & & & & & & & & & & & & & & & & \\
& & & & & & -\frac{1}{2} & 1 & & & & & & & & & & & & & & & & \\
& & & & & & -\frac{1}{2} & & 1 & & & & & & & & & & & & & & & \\
& & & & & & & -\frac{1}{3} & & 1 & & & & & & & & & & & & & & \\
& & & & & & & -\frac{1}{3} & & & 1 & & & & & & & & & & & & & \\
& & & & & & & -\frac{1}{3} & -\frac{1}{2} & & & 1 & & & & & & & & & & & & \\
& & & & & & & & -\frac{1}{2} & & & & 1 & & & & & & & & & & & \\
& & & & & & & & & & & & & 1 & & -\frac{1}{2} & & & & & & & & \\
& & & & & & & & & & & & & -1 & 1 & & & & & & & & & \\
& & & & & & & & & & & & & & -\frac{1}{2} & 1 & & & & & & & & \\
& & & & & & & & & & & & & & -\frac{1}{2} & & 1 & & & & & & & \\
& & & & & & & & & & & & & & & -\frac{1}{2} & & 1 & & & & & & \\
& & & & & & & & & & & & & & & & & & 1 & -\frac{1}{5} & -\frac{1}{5} & -\frac{1}{5} & & \\
& & & & & & & & & & & & & & & & & & -\frac{1}{5} & 1 & -\frac{1}{5} & -\frac{1}{5} & & \\
& & & & & & & & & & & & & & & & & & -\frac{1}{5} & -\frac{1}{5} & 1 & -\frac{1}{5} & & \\
& & & & & & & & & & & & & & & & & & -\frac{1}{5} & -\frac{1}{5} & -\frac{1}{5} & 1 & & \\
& & & & & & & & & & & & & & & & & & -\frac{1}{5} & -\frac{1}{5} & -\frac{1}{5} & -\frac{1}{5} & 1 & \\
& & & & & & & & & & & & & & & & & & -\frac{1}{5} & -\frac{1}{5} & -\frac{1}{5} & -\frac{1}{5} & & 1 \\
\end{bmatrix}
}^\matr{B}
\cdot
\overbrace{
\begin{pmatrix} 
A \\
C \\
D \\
E \\
F \\
G \\
H \\
I \\
J \\
K \\
L \\
M \\
N \\
O \\
P \\
Q \\
R \\
S \\
T \\
U \\
V \\
W \\
X \\
Y \\
\end{pmatrix}
}^\matr{S}
\label{eq:problem}
\end{equation}%
}
\end{sideways}


\begin{sideways}
\parbox{\textheight}{%
\setlength\arraycolsep{4pt}
\setcounter{MaxMatrixCols}{24}
\begin{equation}
\overbrace{
\begin{pmatrix} 
A \\
C \\
D \\
E \\
F \\
G \\
H \\
I \\
J \\
K \\
L \\
M \\
N \\
O \\
P \\
Q \\
R \\
S \\
T \\
U \\
V \\
W \\
X \\
Y \\
\end{pmatrix}
}^\matr{S}
=
\begin{pmatrix} 
1 \\
1 \\
2 \\
1 \\
1 \\
1 \\
1 \\
1.5 \\
1.5 \\
1.5 \\
1.5 \\
2.25 \\
1.75 \\
2.333333 \\
3.333333 \\
2.666667 \\
2.666667 \\
2.333333 \\
2.5 \\
2.5 \\
2.5 \\
2.5 \\
3 \\
3 \\
\end{pmatrix}
=
\overbrace{
\begin{bmatrix} 
1 &  &  &  &  &  &  &  &  &  &  &  &  &  &  &  &  &  &  &  &  &  &  &  \\
 & 1 &  &  &  &  &  &  &  &  &  &  &  &  &  &  &  &  &  &  &  &  &  &  \\
 & 1 & 1 &  &  &  &  &  &  &  &  &  &  &  &  &  &  &  &  &  &  &  &  &  \\
 &  &  & 1 &  &  &  &  &  &  &  &  &  &  &  &  &  &  &  &  &  &  &  &  \\
 &  &  &  & 1 &  &  &  &  &  &  &  &  &  &  &  &  &  &  &  &  &  &  &  \\
 &  &  &  &  & 1 &  &  &  &  &  &  &  &  &  &  &  &  &  &  &  &  &  &  \\
 &  &  &  &  &  & 1 &  &  &  &  &  &  &  &  &  &  &  &  &  &  &  &  &  \\
 &  &  &  &  &  & 0.500 & 1 &  &  &  &  &  &  &  &  &  &  &  &  &  &  &  &  \\
 &  &  &  &  &  & 0.500 &  & 1 &  &  &  &  &  &  &  &  &  &  &  &  &  &  &  \\
 &  &  &  &  &  & 0.167 & 0.333 &  & 1 &  &  &  &  &  &  &  &  &  &  &  &  &  &  \\
 &  &  &  &  &  & 0.167 & 0.333 &  &  & 1 &  &  &  &  &  &  &  &  &  &  &  &  &  \\
 &  &  &  &  &  & 0.417 & 0.333 & 0.500 &  &  & 1 &  &  &  &  &  &  &  &  &  &  &  &  \\
 &  &  &  &  &  & 0.250 &  & 0.500 &  &  &  & 1 &  &  &  &  &  &  &  &  &  &  &  \\
 &  &  &  &  &  &  &  &  &  &  &  &  & 1.333 & 0.333 & 0.667 &  &  &  &  &  &  &  &  \\
 &  &  &  &  &  &  &  &  &  &  &  &  & 1.333 & 1.333 & 0.667 &  &  &  &  &  &  &  &  \\
 &  &  &  &  &  &  &  &  &  &  &  &  & 0.667 & 0.667 & 1.333 &  &  &  &  &  &  &  &  \\
 &  &  &  &  &  &  &  &  &  &  &  &  & 0.667 & 0.667 & 0.333 & 1 &  &  &  &  &  &  &  \\
 &  &  &  &  &  &  &  &  &  &  &  &  & 0.333 & 0.333 & 0.667 &  & 1 &  &  &  &  &  &  \\
 &  &  &  &  &  &  &  &  &  &  &  &  &  &  &  &  &  & 1.250 & 0.417 & 0.417 & 0.417 &  &  \\
 &  &  &  &  &  &  &  &  &  &  &  &  &  &  &  &  &  & 0.417 & 1.250 & 0.417 & 0.417 &  &  \\
 &  &  &  &  &  &  &  &  &  &  &  &  &  &  &  &  &  & 0.417 & 0.417 & 1.250 & 0.417 &  &  \\
 &  &  &  &  &  &  &  &  &  &  &  &  &  &  &  &  &  & 0.417 & 0.417 & 0.417 & 1.250 &  &  \\
 &  &  &  &  &  &  &  &  &  &  &  &  &  &  &  &  &  & 0.500 & 0.500 & 0.500 & 0.500 & 1 &  \\
 &  &  &  &  &  &  &  &  &  &  &  &  &  &  &  &  &  & 0.500 & 0.500 & 0.500 & 0.500 &  & 1 \\
\end{bmatrix}
}^{\matr{B}^{-1}}
\cdot
\overbrace{
\begin{pmatrix} 
1 \\
1 \\
1 \\
1 \\
1 \\
1 \\
1 \\
1 \\
1 \\
1 \\
1 \\
1 \\
1 \\
1 \\
1 \\
1 \\
1 \\
1 \\
1 \\
1 \\
1 \\
1 \\
1 \\
1 \\
\end{pmatrix}
}^{\matr{J}}
\label{eq:solution}
\end{equation}%
}
\end{sideways}

\end{document}